\documentclass[twocolumn]{aastex631}
\hypersetup{colorlinks,linkcolor={cyan},citecolor={cyan},urlcolor={cyan}} 

\usepackage{lineno}
\usepackage[T1]{fontenc}

\usepackage{color}

\def\ergs{\textit{$\rm erg\ s^{-1}$}}

\def\lbol{\textit{$L_{\rm bol}$}}
\def\re{\textit{$R_{\rm e}$}}
\def\mbulge{\textit{$M_{\rm bulge}$}}
\def\msun{\textit{$M_{\odot}$}}
\def\ss{\textit{$\rm S\acute{e}rsic$}}
\def\hb{\textit{$\rm H\beta$}}
\def\oiii{[\textsc{O$\,$iii}]}
\def\chisq{\textit{$\chi_{\nu}^2$}}
\def\mbh{\textit{$M_{\rm BH}$}}

\def\fagn{\textit{$f_{\rm agn}$}}

\def\kms{\textit{$\rm km\ s^{-1}$}}
\def\ergs{\textit{$\rm erg\ s^{-1}$}}

\begin{document}

\title{JWST Confirms the Nature of CID-42}

\author[0000-0002-1605-915X]{Junyao Li}
\affiliation{Department of Astronomy, University of Illinois at Urbana-Champaign, Urbana, IL 61801, USA}
\correspondingauthor{Junyao Li}
\email{junyaoli@illinois.edu}

\author[0000-0001-5105-2837]{Mingyang Zhuang}
\affiliation{Department of Astronomy, University of Illinois at Urbana-Champaign, Urbana, IL 61801, USA}

\author[0000-0003-1659-7035]{Yue Shen}
\affiliation{Department of Astronomy, University of Illinois at Urbana-Champaign, Urbana, IL 61801, USA}
\affiliation{National Center for Supercomputing Applications, University of Illinois at Urbana-Champaign, Urbana, IL 61801, USA}

\begin{abstract}

The galaxy CID-42 (CXOC J100043.1+020637.2) at $z=0.359$ has been proposed to contain a promising candidate for a gravitational wave recoiling supermassive black hole (SMBH), a slingshot SMBH from a triple-SMBH interaction, or a kpc-scale dual Active Galactic Nuclei (AGN). These claims were primarily based on a pair of bright cores separated by $\sim 0\farcs5$ resolved in optical HST imaging. Existing HST, Chandra and ground-based imaging and spectroscopy are unable to confirm either scenario. With improved spatial resolution, depth, and IR wavelength coverage, NIRCam multi-band imaging from the COSMOS-Web JWST treasury program well resolved the two cores in CID-42, revealing a significant stellar bulge for both cores (with stellar masses of $\sim 10^{10}\,M_\odot$ for both). 
JWST imaging further revealed that only the SE core contains an unobscured AGN point source, based on both image decomposition and spectral energy distribution fitting. There is no evidence for AGN activity in the NW core. These new observations unambiguously rule out the GW-recoiling and slingshot SMBH scenarios, and establish CID-42 as a low-redshift merging pair of galaxies, with only one active AGN in the system. These results demonstrate the unparalleled capabilities of JWST (even with imaging alone) in studying the galactic-scale environment of merging galaxies and SMBHs.
\end{abstract}
%\keywords{Active galactic nuclei (16); Galaxy interactions (600); Double quasars (406) }

\section{Introduction}
Galaxy mergers represent a fundamental process within the hierarchical structure formation paradigm. This process is widely recognized to play a critical role in driving transitions in  galaxy structures, triggering star formation, and fueling the growth of SMBHs \citep[e.g.,][]{Hopkins2008, Conselice2014}. When two galaxies merge, the SMBHs at the centers of each galaxy can be simultaneously activated and manifest as a dual AGN \citep[e.g.,][]{DeRosa2019}. During the late merging phase, the two SMBHs form a close binary system and gradually shrink their orbit under the influence of dynamical friction via gas dynamics and stellar interactions \citep{Begelman1980}. The final coalescence process is characterized by the emission of strong gravitational waves (GWs), which carries away energy as well as linear and angular momentum from the binary system that facilitates the ultimate merge of the BHs \citep[e.g.,][]{Hughes2002}. If the GWs are emitted anisotropically (depends on BH spin, mass ratio, spin-orbit orientation), a gravitational ``kick" can occur to conserve the linear momentum. This kick can be powerful enough to eject the merged SMBH from galactic center, along with its accretion disk, broad-line region, and the nuclear star cluster bound to it, with a recoil velocity up to $\sim4000\ \kms$ \citep[e.g.,][]{Loeb2007, Campanelli2007, Merritt2009}. This process could produce an AGN with significant spatial and/or kinematic offset signatures \citep[e.g.,][]{Bonning2007, Blecha2008}.

When dynamical friction alone is insufficient to propel the SMBH binary into the GW-dominated regime, the binary stalls until the involvement of a third galaxy/SMBH. The subsequent interaction among the triple-SMBH system can further harden the binary system, leading to its eventual coalescence and the kick out of one of the three SMBHs, a phenomenon referred to as the slingshot ejection \citep[e.g.,][]{Hoffman2007}.

The presence of recoiling BHs provides indirect evidence of GW emission from SMBH mergers. Observationally confirming the recoiling or the slingshot scenarios could have a tremendous impact on our understanding of the growth history of massive BHs \citep[e.g.,][]{Volonteri2003}. However, despite the continued observational effort of identifying promising candidates, none of them has been confirmed thus far \citep[e.g.,][]{Morishita2022}.

In this paper, we focus on CID-42 (CXOC J100043.1+020637.2), which has been the subject of intense multiwavelength monitoring due to its exceptional properties that make it one of the most promising candidates for a recoiling SMBH \citep[][]{Comerford2009, Civano2010, Civano2012, Blecha2013, Novak2015}. It is a merging galaxy located at $z=0.359$ and exhibits a prominent tidal tail. Two bright nuclei, NW and SE, are well resolved in HST $I$-band (F814W) imaging. Notably, the SE nucleus is offset from the galaxy center (presumably at the location of the NW nucleus; \citealt{Civano2010}) by $\sim 0\farcs495$, corresponding to $\sim2.46$~kpc at its redshift. Chandra X-ray observations further reveal that only the SE nucleus is detected in X-rays, which is responsible for producing the broad $\rm H\beta$ line seen in the spatially-unresolved, ground-based spectroscopy. Interestingly, the broad $\rm H\beta$ line shows a significant velocity offset of $\sim1300\,\kms$ from its narrow component. 

Three competing scenarios have been proposed to explain the unusual properties of CID-42. (1) a GW recoiling SMBH \citep{Civano2010}: NW is a compact stellar bulge that defines the center of the galaxy, while SE is a broad-line AGN being kicked off from the galaxy center  by GW recoiling. (2) A type 1 (unobscured) AGN recoiling from a type 2 (obscured) AGN in a slingshot \citep{Civano2010}: NW might be an obscured AGN resulting from the coalescence of a SMBH binary that is responsible for the narrow \hb~line (NL) {and contribute partly to the strong Fe~K$\alpha$ line ($\rm EW\sim500$~eV) seen in the X-ray spectrum}, while SE is the unobscured slingshot-ejected AGN producing the kinematically-offset broad \hb~line (BL). (3) A kpc-scale dual AGN \citep{Comerford2009}: both SE and NW are inspiraling AGNs within the galaxy merger, producing a double-peak \oiii~line that places both nuclei in the AGN-dominated region on the BPT diagram \citep{Kewley2006}. 

However, there are significant caveats associated with all these interpretations. The recoiling and slingshot scenarios are built upon the spatial and velocity offsets of the system. \cite{Civano2010} argued that a BL/NL offset of $\sim1300\ \kms$ is too large to be explained by the broad-line region kinematics or orbital motion of kpc-scale dual AGNs. However, it is not unusual to observe such a large velocity offset for the broad \hb~line even in single quasars \citep{Shen2016}. In fact, about $5\%$ of the SDSS DR16 quasars with sufficient $\rm S/N>10$ per pixel \citep{Wu2022} have the peak of the broad \hb~line shifted by more than $1000\ \kms$ from that of its entire line profile (presumably tracing the peak of its NL component). Regarding the spatial offset, the SE nucleus could be associated with its own bulge that is not readily resolvable by optical HST imaging. The double-peak \oiii~line proposed in \cite{Comerford2009} is also elusive and could be affected by the low S/N of the spectra; in addition, such NL profiles are more likely caused by narrow-line-region kinematics rather than dual AGNs \citep[e.g.,][]{Shen2011b}. 

With the advent of JWST, we now have the opportunity to study CID-42 with unprecedented spatial resolution with NIRCam imaging \citep{Rieke2023} across a wide range of IR wavelengths. We perform two-dimensional AGN-galaxy image decomposition and construct spatially-resolved spectral energy distribution (SED) to unveil the nature of the two bright nuclei, the underlying host galaxy, and their spatial offset. Throughout this paper we adopt a flat $\Lambda$CDM cosmology with $\Omega_{\Lambda}=0.7$ and $H_{0}=70\,{\rm km\,s^{-1}Mpc^{-1}}$. Magnitudes are given in the AB system.

\begin{figure*}
\centering
\includegraphics[width=0.7\linewidth]{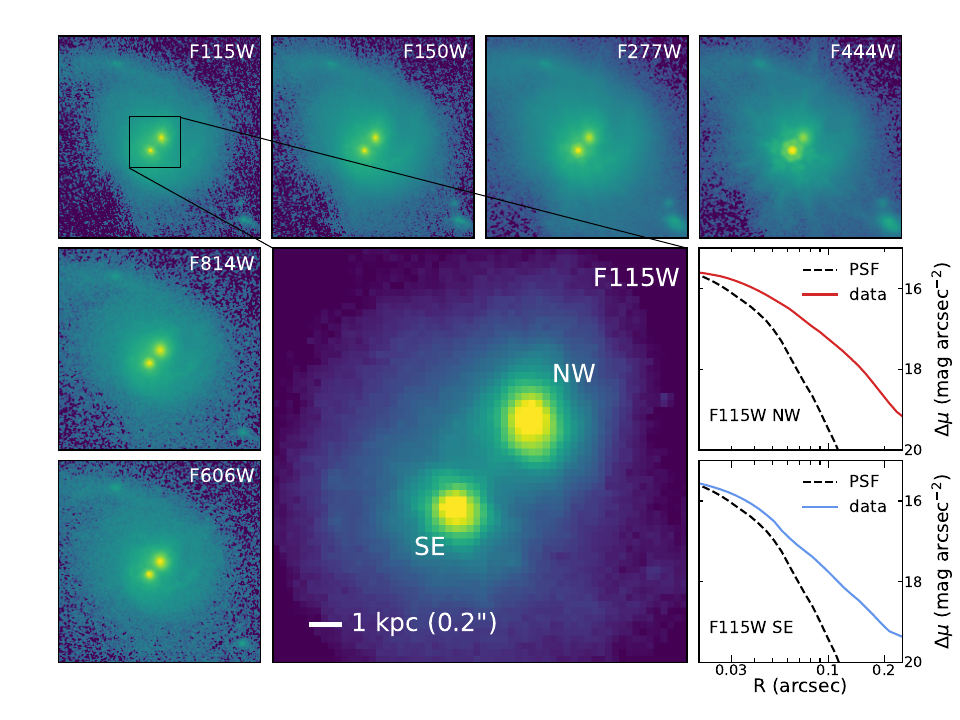}
\caption{Multi-band HST and JWST imaging data for CID-42. The bottom middle and right panels display the zoomed-in view of the two cores in the F115W band and their corresponding surface brightness profiles compared with that of the PSF (normalized at the peak value).}
\label{fig:image}
\end{figure*}

\section{Data}
COSMOS-Web \citep[PIs: Kartaltepe \& Casey, ID=1727]{Casey2022} is a 255 hour treasury imaging survey in JWST Cycle 1. It maps a contiguous 0.54 deg$^2$ region in four NIRCam filters (F115W, F150W, F277W, and F444W) that reaches a $5\sigma$ point source depth of $27.5-28.2$ magnitudes. In parallel, MIRI imaging \citep{Bouchet2015} over a 0.19 deg$^2$ region is taken in the F770W filter. 

CID-42 falls within the footprint of visit CWEBTILE-4-11 with observation number 99. Currently, only NIRCam data are available. The MIRI observations covering CID-42 have been scheduled in December 2023. We retrieve uncalibrated raw NIRCam data (\_uncal.fits) from MAST and use the \texttt{jwst} pipeline (version 1.10.2) with the Calibration Reference Data System (CRDS) version of 11.17.0 (context file \texttt{jwst\_1089.pmap}) and custom steps for NIRCam image data reduction. Details of the reduction are provided in Zhuang et al. (in prep.). We briefly summarize our steps below. We apply detector-level corrections to the uncalibrated images using stage~1 of the \texttt{jwst} pipeline and 1/$f$ noise correction using an adapted script from the CEERS team \citep{Bagley+2023ApJ}. The output count-rate images from the previous step are processed with stage 2 of the \texttt{jwst} pipeline to produce fully calibrated images. We then perform 2-dimensional background subtraction to the individual exposures using SExtractor \citep{SExtractor}. Finally, we apply stage 3 of the \texttt{jwst} pipeline to align astrometry with the COSMOS2020 catalog \citep{COSMOS2020}, perform outlier rejection, and resample the individual exposures to produce the final combined images. The output mosaics are drizzled to a pixel scale of 0\farcs03 pixel$^{-1}$ with \texttt{pixfrac=0.8}. 

We use the software \texttt{PSFEx} \citep{PSFEx} to construct point spread function (PSF) models, which has been demonstrated to provide the best PSF models among commonly used methods \citep{Zhuang&Shen2023}. We follow the procedures in \cite{Zhuang&Shen2023} and construct PSF models by using all the high signal-to-noise ratio ($>100$) point-like sources in the mosaics. 

We also complement the NIRCam imaging data with fully reduced HST/ACS imaging (0\farcs03 pixel$^{-1}$; same as NIRCam data) in the F606W and F814W bands from the CANDELS survey \citep{Koekemoer2011} using
the publicly released v1.0 mosaics. The effective PSF model \citep{Anderson2000} for HST imaging is built from unsaturated point-like sources in the ACS cutout ($2^\prime \times 2^\prime$) using the {\tt{EPSFBuilder}} method available in the python package {\tt{photutils}} \citep{Bradley2022}.

\section{Imaging Analyses}
\label{sec:method}

\begin{figure*}
\centering
\includegraphics[width=0.75\linewidth]{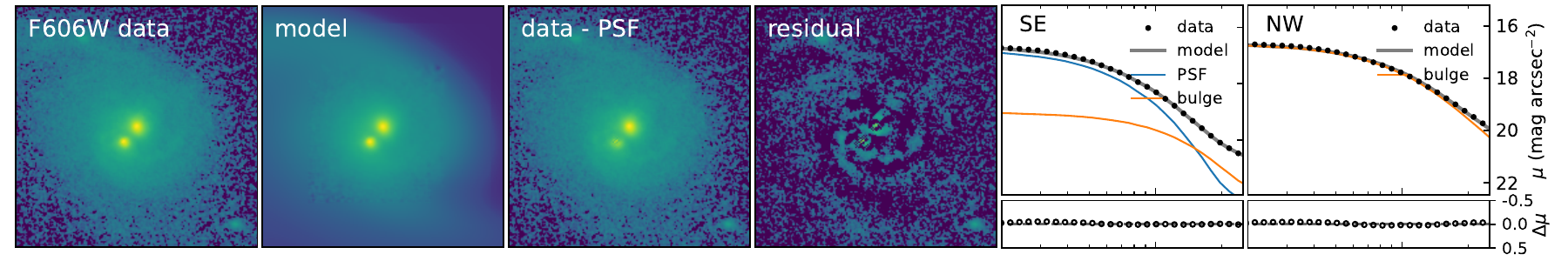}
\includegraphics[width=0.75\linewidth]{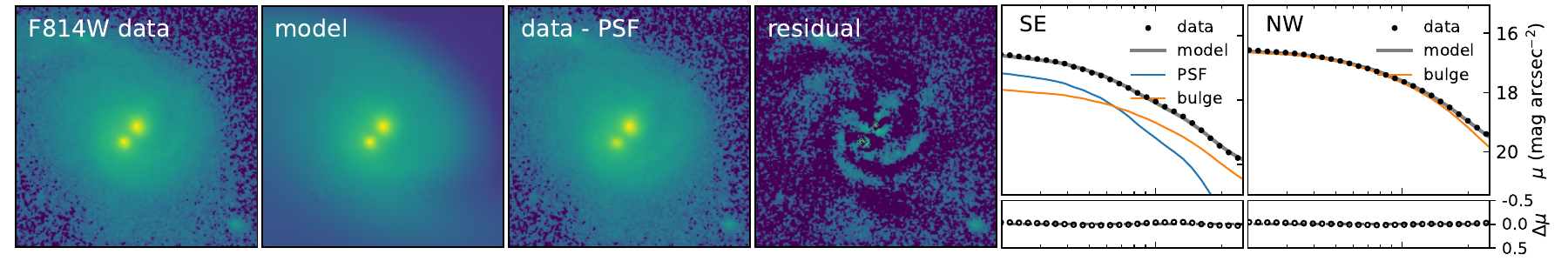}
\includegraphics[width=0.75\linewidth]{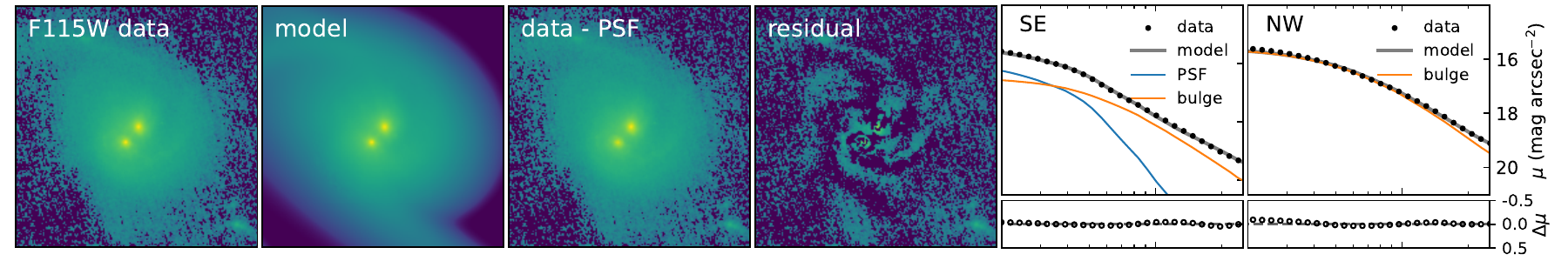}
\includegraphics[width=0.75\linewidth]{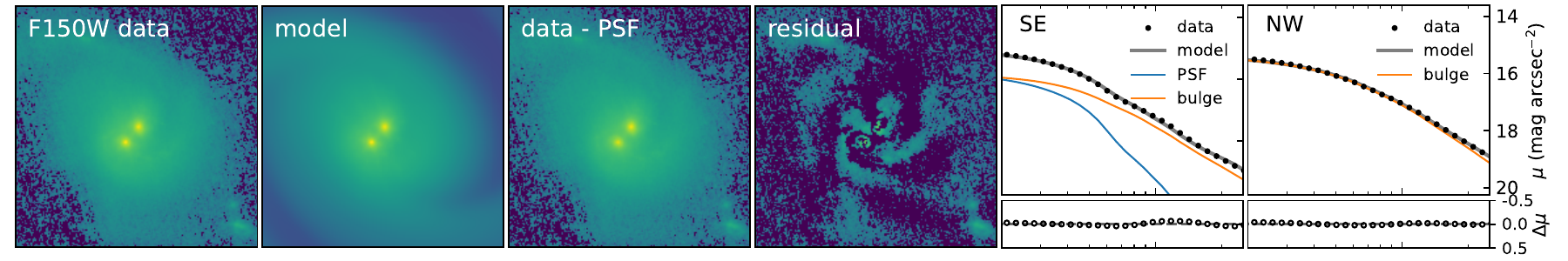}
\includegraphics[width=0.75\linewidth]{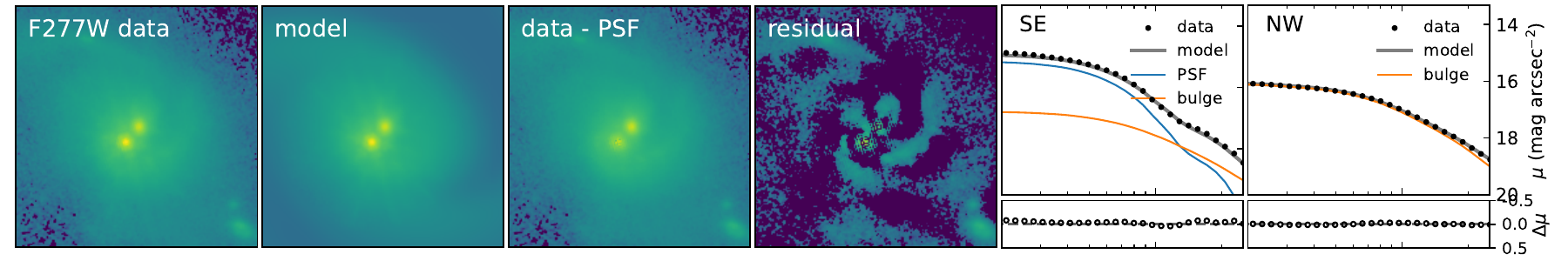}
\includegraphics[width=0.75\linewidth]{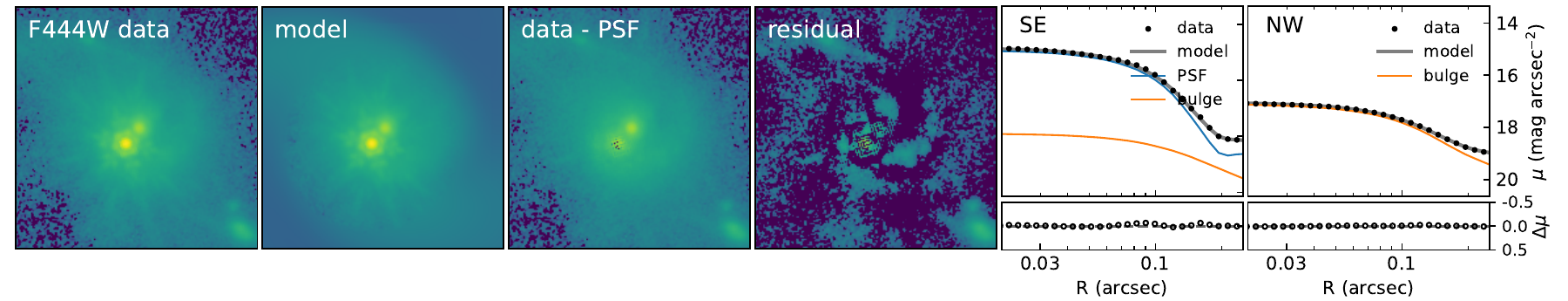}
\caption{Image decomposition results in the six-band HST and JWST images using Model A. The panels are (1) observed data, (2) best-fit model, (3) PSF-subtracted data (i.e., galaxy emission only), (4) residual map, (5) and (6) radial surface brightness profiles of the SE and NW cores within the central $\sim1$~kpc region, with the bottom panels showing the difference ($\Delta \mu$) between data and model. The low-surface-brightness \ss~components are not shown in the surface brightness plot since they are much fainter than the PSF and bulge components.}
\label{fig:image_decomp}
\end{figure*}

Figure \ref{fig:image} displays the multi-band HST and JWST images of CID-42. After conducting an initial visual inspection of the JWST data, it becomes evident that the two bright cores are not point-like. Instead, the central peak emission of each core exhibits an elongated morphology, suggestive of an underlying galaxy component, possibly a stellar bulge associated with each core. In fact, the surface brightness decomposition of the HST F814W image in \cite{Civano2010} already revealed that the NW nucleus is best described by a compact bulge rather than a point source. The extended galaxy features are most conspicuously observed in the F115W (a zoomed-in view of the central region is shown in Figure \ref{fig:image}) and the F150W bands where the resolution is highest and the host contribution is more pronounced than those in the HST bands. Moreover, in Figure \ref{fig:image}, we present the radial surface brightness profile of each nucleus within the central 1 kpc-radius region ($0\farcs2$) in F115W, compared with that of the PSF model (normalized at the peak value). The surface brightness profiles of both cores deviate significantly from that of the PSF, indicating the presence of a bulge associated with each core. 

The existence of a stellar bulge associated with the SE nucleus (i.e., the proposed kicked-off SMBH) would significantly impact the interpretation of the CID-42 system. For example, if the bulge is massive enough, then the recoiling SMBH and the slingshot ejection scenarios would be ruled out. In the following sections, we conduct a more quantitative analysis to confirm the presence of bulges, measure their stellar masses, and explore the potential presence of an AGN in the NW core. The SE nucleus contains an unobscured broad-line AGN as confirmed in previous X-ray and optical observations, which we will verify with our JWST+HST imaging analysis. 

\begin{figure*}
\centering
\includegraphics[width=0.75\linewidth]{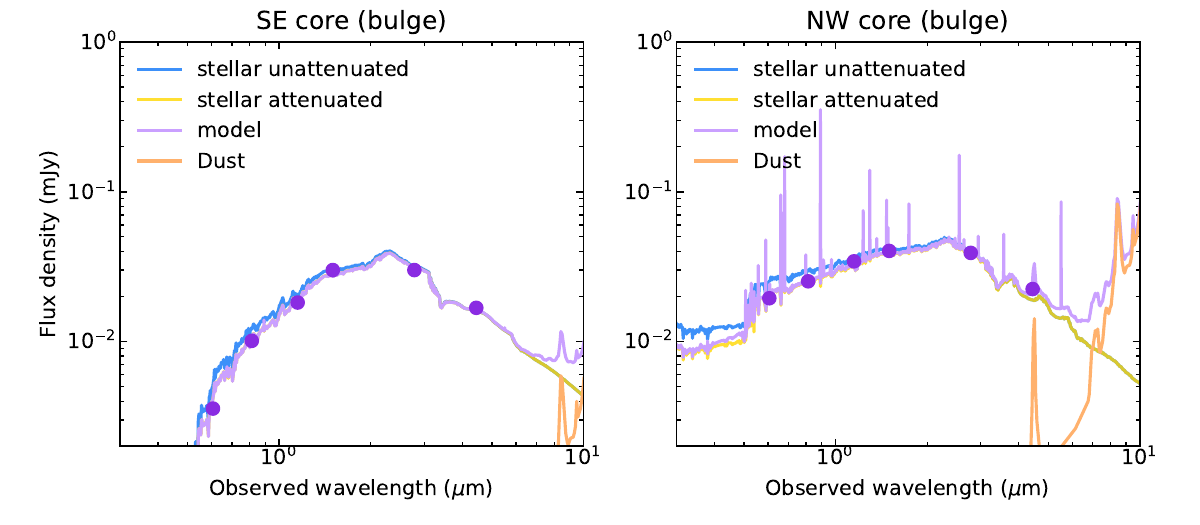}
\caption{SED fitting to the decomposed bulge flux within a 1 kpc-radius aperture for the SE and NW nuclei, respectively. The observed flux (purple points) has been subtracted for AGN contribution through image decomposition. The best-fit total model is shown in the purple curve and its components (stellar unattenuated, stellar attenuated, and dust emission) are also displayed.}
\label{fig:bulge_sed}
\end{figure*}

\subsection{Image decomposition}
\label{subsec:image}
We model the surface brightness of CID-42 with GALFIT \citep{Peng2010} to disentangle the extended galaxy emission from the point-like AGN component. Motivated by the initial image inspection above, our surface brightness model is defined as follows. 1) SE nucleus: we utilize a combined PSF + \ss~profile to fit the broad-line AGN and its underlying bulge\footnote{For convenience, we refer to this compact stellar component revealed by the subsequent GALFIT fitting results as a bulge.} component. The centers of the PSF and the \ss~components are free to vary. 2) NW nucleus: by default, we adopt a \ss~profile to fit the compact stellar bulge \citep{Civano2010}. Additionally, we explore the potential presence of a low luminosity and/or obscured AGN by adding a PSF component and evaluating the improvement in the goodness of the fit. 3) {Low-surface-brightness (LSB) components: we incorporate an exponential disk profile (i.e., \ss~index $n=1$) for each nucleus}, modulated by the $m=1$ Fourier mode and power-law coordinate rotation, with the same centroid as the \ss~profile for each bulge, to account for the extended asymmetric stellar disk emission and tidal tails that cannot be adequately modeled by the smooth \ss~profiles. Models with and without an additional PSF component for the NW nucleus are referred to as Models A and B, respectively. Objects unassociated with CID-42 in the field of view are masked during the fitting process.

\begin{table*}
\centering
\renewcommand{\arraystretch}{1.1}
\caption{GALFIT surface brightness fitting result using Model A. Given the large number of model parameters, we only show the most relevant parameters, namely the integrated model magnitude, half-light radius (\re), and \ss~index ($n$).}
\begin{tabular}{ccccccc}
\hline
\hline
Parameters & F606W & F814W & F115W & F150W & F277W & F444W \\
\hline
\multicolumn{7}{c}{SE AGN (PSF)} \\
mag & $21.15_{-0.01}^{+0.01}$ & $21.35_{-0.01}^{+0.01}$ & $21.28_{-0.01}^{+0.01}$ & $20.92_{-0.01}^{+0.01}$ & $19.16_{-0.01}^{+0.01}$ & $18.43_{-0.01}^{+0.01}$\\
\hline
\multicolumn{7}{c}{SE bulge (\ss)} \\
mag & $22.10_{-0.02}^{+0.02}$ & $20.78_{-0.01}^{+0.01}$ & $20.19_{-0.01}^{+0.01}$ & $19.24_{-0.01}^{+0.01}$ & $19.25_{-0.02}^{+0.02}$ & $19.91_{-0.02}^{+0.02}$\\
$R_{\rm e}$ (kpc) & $0.44_{-0.01}^{+0.01}$ & $0.43_{-0.01}^{+0.01}$ & $0.44_{-0.01}^{+0.01}$ & $0.81_{-0.01}^{+0.01}$ & $0.93_{-0.01}^{+0.01}$ & $0.97_{-0.01}^{+0.01}$\\
$n$ & $0.75_{-0.05}^{+0.05}$ & $3.01_{-0.08}^{+0.08}$ & $2.37_{-0.04}^{+0.04}$ & $4.94_{-0.11}^{+0.11}$ & $2.96_{-0.11}^{+0.11}$ & $2.02_{-0.07}^{+0.07}$\\
\hline
\multicolumn{7}{c}{NW bulge (\ss)} \\
mag & $20.13_{-0.01}^{+0.01}$ & $19.86_{-0.01}^{+0.01}$ & $19.49_{-0.01}^{+0.01}$ & $19.23_{-0.01}^{+0.01}$ & $19.24_{-0.01}^{+0.01}$ & $19.89_{-0.01}^{+0.01}$\\
$R_{\rm e}$ (kpc) & $0.35_{-0.01}^{+0.01}$ & $0.38_{-0.01}^{+0.01}$ & $0.41_{-0.01}^{+0.01}$ & $0.42_{-0.01}^{+0.01}$ & $0.41_{-0.01}^{+0.01}$ & $0.44_{-0.01}^{+0.01}$\\
$n$ & $2.10_{-0.02}^{+0.02}$ & $2.06_{-0.02}^{+0.02}$ & $2.47_{-0.02}^{+0.02}$ & $2.66_{-0.02}^{+0.02}$ & $2.38_{-0.02}^{+0.02}$ & $2.57_{-0.04}^{+0.04}$\\
  \hline
 \multicolumn{7}{c}{SE LSB (exponential disk + $m=1$ Fourier mode + power-law rotation)} \\
mag & $20.15_{-0.01}^{+0.01}$ & $19.62_{-0.01}^{+0.01}$ & $19.55_{-0.01}^{+0.01}$ & $19.06_{-0.01}^{+0.01}$ & $18.90_{-0.01}^{+0.01}$ & $19.65_{-0.01}^{+0.01}$\\
$R_{\rm e}$ (kpc) & $6.90_{-0.07}^{+0.07}$ & $5.37_{-0.04}^{+0.04}$ & $2.42_{-0.01}^{+0.01}$ & $2.88_{-0.02}^{+0.02}$ & $3.17_{-0.02}^{+0.02}$ & $2.95_{-0.04}^{+0.04}$\\
 \hline
 \multicolumn{7}{c}{NW LSB (exponential disk + $m=1$ Fourier mode + power-law rotation)} \\
mag & $20.82_{-0.01}^{+0.01}$ & $20.17_{-0.01}^{+0.01}$ & $19.51_{-0.01}^{+0.01}$ & $19.67_{-0.01}^{+0.01}$ & $19.56_{-0.01}^{+0.01}$ & $19.74_{-0.01}^{+0.01}$\\
$R_{\rm e}$ (kpc) & $2.64_{-0.03}^{+0.03}$ & $2.59_{-0.02}^{+0.02}$ & $4.36_{-0.05}^{+0.05}$ & $2.21_{-0.02}^{+0.02}$ & $2.35_{-0.02}^{+0.02}$ & $3.15_{-0.02}^{+0.02}$\\
\hline
\multicolumn{7}{c}{Goodness of fit} \\
\chisq & 1.77 & 2.14 & 1.10 & 1.51 & 9.38 & 6.71\\
\hline\\
\end{tabular}
\label{table:galfit}
\end{table*}

Figure \ref{fig:image_decomp} and {Table \ref{table:galfit}} summarize the decomposition results in the six-band HST and JWST images using Model A. By comparing the reduced \chisq~values of  Models A and B, we find that adding an extra PSF profile for the NW nucleus has a negligible impact on the fitting goodness in all the HST and JWST bands. The NW nucleus can always be well fitted by the \ss~profile alone as shown in Figure~\ref{fig:image_decomp}. It is (or dominated by) a compact stellar bulge with a half-light radius of $\re\sim0.41_{-0.04}^{+0.02}$~kpc and a \ss~index of $n\sim2.45_{-0.37}^{+0.17}$, where the best-fit parameters and their uncertainties are given based on the 16th, 50th, and 84th percentiles of the \ss~parameters obtained in each band. The fact that the long wavelength filter F444W, which traces hot dust emission, also lacks a significant point source component, demonstrates that the non-detection of an AGN in the short-wavelength bands is not caused by dust extinction. Instead, it indicates that the AGN activity in the NW core is either intrinsically weak or lack thereof. 

The presence of a prominent bulge for the SE nucleus is evident in the PSF-subtracted images across all filters. The F444W emission for the SE nucleus is dominated by the AGN point source (from the dusty torus), while the AGN and the bulge contribute comparable fluxes in shorter-wavelength bands. The bulge of the SE nucleus is also compact with $\re\sim0.63_{-0.18}^{+0.32}$~kpc and $n\sim2.65_{-0.90}^{+0.73}$. Notably, its position is well aligned with that of the best-fit point source component, with a spatial offset of $\lesssim 1$~pixel {($0\farcs03$) in all bands}.

{Next, we measure the bulge flux for each nucleus in multiple bands from the PSF-subtracted images within a 1 kpc-radius aperture centered on each nucleus. The resolution of each image is matched to that of the lowest resolution image (i.e., F444W) using a matching kernel derived from Fourier transform \citep{Aniano2011} available in {\tt{photutils}}.
When measuring the aperture photometry for the NW nucleus, the PSF and \ss~components centered on the SE nucleus were subtracted, and vice versa, to avoid cross contamination. We then employ SED fitting with CIGALE v2022.1 \citep{Boquien2019, Yang2020} to estimate their stellar masses. We assume a \cite{BC03} stellar population model with solar metallicity, a \cite{Chabrier2003} initial mass function,  a \cite{Calzetti2000} extinction law with a color excess $E(B-V)_{\rm lines}$ of 0.0 -- 0.6 mag, a \cite{Dale2014} dust emission model ($\alpha=$ 1.5, 2.0, 2.5), and include nebular lines. We adopt a delayed star formation history (SFH) with a stellar age of 1.0 -- 9.0 Gyr and an e-folding time of the main stellar population of 0.1 -- 10.0 Gyr. We also allow for a recent burst with a mass fraction up to $10\%$ to account for a potential starburst triggered by mergers. In the fitting, we set the {\tt{additionalerror}} parameter (i.e., relative error added in quadrature to the flux uncertainties) in CIGALE to $2\%$ to account for typical uncertainties of the JWST absolute flux calibration \citep[e.g.,][]{Bagley+2023ApJ}.}

Figure~\ref{fig:bulge_sed} presents the constructed bulge SEDs (corrected for galactic extinction) and the results of the model fitting. Both bulges are dominated by an old stellar population with an age of $\sim9$~Gyr for SE and $\sim4$~Gyr for NW. The NW nucleus has a more recent, smoothly declined SFH with an e-folding time of $\sim 3.5$~Gyr. The star formation process in the SE nucleus is characterized by an e-folding time of $\sim0.6$~Gyr and has almostly ceased within the past $\sim5$~Gyr. The derived stellar masses for the SE and NW nuclei are $1.5\pm 0.12 \times 10^{10}\,\msun$ and $5.8\pm 0.5 \times 10^{9}\,\msun$, respectively. The systematic uncertainties in the stellar mass estimates due to model assumptions and the adoption of different SED fitting codes are unlikely to exceed a factor of few \citep[e.g.,][]{Pacifici2023}.

The fact that the SE nucleus is well-centered on its own bulge, which has a stellar mass of $\mbulge\sim10^{10}\,\msun$, unambiguously rules out the recoiling SMBH or slingshot scenarios. In either case, {a bulge this massive cannot be bound to the kicked-off SMBH}. According to \cite{Merritt2009}, the total mass in stars ($M_b$) that will remain bound to the BH after the kick can be estimated as 
\begin{equation}
\frac{M_{\rm b}}{\mbh} = 2\times 10^{-4} \left(\frac{\mbh}{10^7\,\msun} \right)^2 \left( \frac{r_\bullet}{10\, \rm pc} \right)^{-2} \left( \frac{V_{\rm k}}{10^3\,\kms} \right)^{-4},
\end{equation}
where the BH mass is estimated to be $\sim 6.5\times 10^7 \,\msun$ based on the broad \hb~line \citep{Civano2010}, $V_{\rm k}$ is the kick off velocity which we take as the BL/NL velocity offset ($\sim1300\,\kms$) as an approximation, and $r_{\bullet}$ is defined as the radius that contains an integrated stellar mass equal to twice \mbh, which is of the same order as the radius of influence of the SMBH ($r_{\rm infl} \equiv G \mbh/\sigma^2 \sim 10\rm \, pc$). The ejected stellar mass is thus on the order of $\sim10^5\,\msun$, which is far less than our measured bulge mass.

\begin{figure*}
\centering
\includegraphics[width=0.75\linewidth]{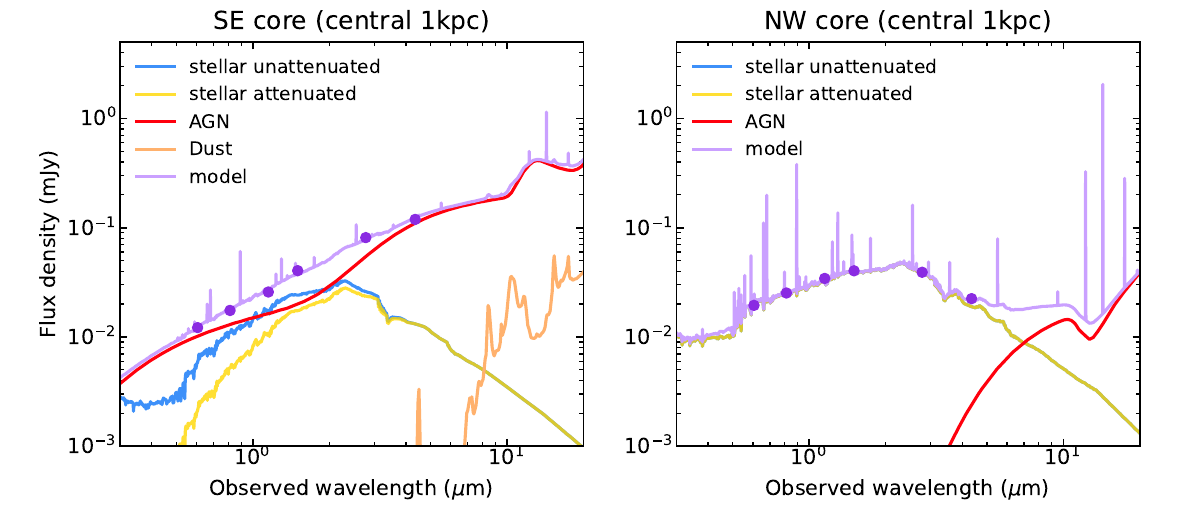}
\caption{SED decomposition result of the two nuclei. The observed flux (purple points) corresponds to the total PSF-matched flux within a 1 kpc-radius aperture centered on each nucleus. The best-fit total model is shown as a purple curve and its components (stellar unattenuated, stellar attenuated, dust, and AGN) are also displayed. {The best-fit dust component for the NW core is zero and thus is not shown in the right panel.}}
\label{fig:sed_decomp}
\end{figure*}

\subsection{SED decomposition for both cores}
In order to gain further insights into the nature of the two cores, we construct spatially resolved (for the SE and NW cores separately) SEDs for the central region of CID-42. SED decomposition is then performed to isolate the galaxy and AGN contribution for each core. Similar to Section \ref{subsec:image}, this is achieved by measuring the PSF-matched flux within a 1~kpc-radius aperture for each band, centered on the SE and NW nuclei. The PSF and \ss~components unrelated with each nucleus were subtracted when measuring the photometry. The resulting SEDs are presented in Figure~\ref{fig:sed_decomp}. It is evident that the SED of the SE nucleus exhibits a steep power-law shape characteristic of a typical type 1 AGN, while the SED of the NW nucleus resembles that of a normal galaxy or a low luminosity/obscured AGN \citep[e.g.,][]{Li2020}.

We then incorporate the SKIRTOR AGN model  \citep{Stalevski2016} with three free parameters (inclination angle, AGN fraction, and $E(B-V) = 0-0.3$ for the polar dust extinction) into the same SED models presented in Section \ref{subsec:image}, and perform a decomposition of the SEDs for the two cores into their AGN and galaxy components. The strength of the AGN component is controlled by the AGN fraction to the total emissions in the F444W band (\fagn). The inclination angle $\theta$ is set to either $30^\circ$ or $70^\circ$ to represent type 1 and type 2 scenarios, respectively.

The resulting best-fit model SEDs are displayed in Figure \ref{fig:sed_decomp}. The derived bulge stellar masses  for the SE and NW nuclei are $1.0\pm 0.2 \times 10^{10}\,\msun$ and $7.1\pm 0.6 \times 10^{9}\,\msun$, respectively, which are in good agreement with those obtained from image decomposition. Consistent with the image decomposition results, the SED of the SE nucleus is dominated by the type 1 AGN ($\theta=30^\circ$) component in the long-wavelength filters, with $\fagn \sim 87\pm2\%$ and $L_{\rm bol} \sim 1.3\times 10^{44}\,\ergs$. 

In contrast, the SED decomposition reveals that the NW nucleus hosts a low-luminosity type 2 AGN ($\theta=70^\circ$), with an observed $\fagn\sim 9.3\pm3.2\%$ and a dust-corrected $\lbol \sim 1.9\times10^{43}\,\ergs$. The detection of a faint obscured AGN by CIGALE is primarily determined by the modest improvement in the reduced $\chi^2$ value when including an AGN component. However, this improvement is statistically insignificant when considering that the number of free parameters also increased, as evaluated through a Bayes Information Criterion test ($\rm \Delta BIC < 2.0$; \citealt{Raftery1995}), and a pure galaxy model can also well explain the SED (Figure \ref{fig:bulge_sed}). Moreover, since only a single F444W band shows a potential excess of an AGN component (Figure \ref{fig:sed_decomp}), and given the complex shape of the AGN and galaxy spectrum in the mid-IR and their degeneracies, it is plausible that our adopted galaxy and AGN templates in CIGALE may not be suitable for CID-42, and we may misidentify the unresolved star formation emission as a low-luminosity AGN. 

{To assess the false positive rate of identifying a weak obscured AGN, we conduct a mock SED test. We generate 1000 mock SEDs of the NW nucleus from the photometry of the best-fit template using only the stellar + dust model in the fitting (purple curve in Figure \ref{fig:bulge_sed}), pertubated by a 2\% random Gaussian flux uncertainty.} We then run CIGALE with AGN models incorporated to examine the best-fit \fagn, which should ideally be close to zero. To simulate the effect of template mismatch, this time we adopt a double exponential SFH, a \cite{Draine2014} dust emission model, and a \cite{Fritz2006} AGN model in the fitting, which are different from those used in decomposing the observed SED and generating the mock SEDs. We find that approximately $37\%$ of the fitted \fagn~values exceed that obtained from decomposing the observed SED ($\sim9.3\%$), indicating that a large fraction of the pure galaxy SED was misidentified as an obscured AGN due to the combined effects of model mismatch, parameter degeneracy, and flux uncertainties. Therefore, we conclude that the presence of a low-luminosity obscured AGN in the NW nucleus is not supported by the data, and a compact stellar bulge is able to explain both the image and SED fitting results (Figures \ref{fig:image_decomp} and \ref{fig:bulge_sed}). 

The combined analysis of imaging decomposition and spatially-resolved SED fitting then suggests that CID-42 is a merging pair of galaxies in an advanced stage of interaction, with only one active AGN in the system. The final coalescence of the two galaxies and their SMBHs has not yet occurred. Subsequent MIRI observations in the F770W band will provide better constraints on the mid-IR SED and the NW nucleus \citep[e.g.,][]{Yang2023}. 

\section{Conclusions}
CID-42, a galaxy merger at $z=0.359$ displaying two closely-separated bright nuclei, a prominent tidal tail, a significant velocity offset between the broad and narrow components of the \hb~line, and X-ray detection for only one of the nuclei, represents one of the most studied candidates for a GW recoiling SMBH. However, alternative scenarios, such as a slingshot AGN or an inspiraling dual AGN system, cannot be ruled out based on past observations.

In this study, leveraging the unparalleled spatial resolution, depth, and IR wavelength coverage offered by the JWST NIRCam imaging, we constrain the nature of this system with comprehensive image decomposition and spatially-resolved SED fitting. 
Our analyses revealed that only one of the bright nuclei, SE, harbors an active type 1 AGN. The other nucleus, NW, is identified as a compact stellar bulge with $\re\sim0.6$~kpc and $\mbulge \sim 10^{10}\,\msun$. 
Importantly, we demonstrated that the AGN point source embedded within the SE core, previously believed to be offset from the galactic center, is actually well-centered on its own bulge with a stellar mass of $\sim10^{10}\,\msun$. These findings unambiguously rule out the recoiling or slingshot SMBH scenarios, establishing CID-42 as a merging pair of galaxies with only one actively accreting SMBH.

Our results demonstrate the exceptional capabilities of JWST in resolving detailed galaxy structures and uncovering embedded AGNs through multi-band imaging and spatially-resolved SED analyses. By extending our case study of CID-42 to a statistical sample of merging galaxies in the COSMOS-Web field, we anticipate improved detection of nuclear activity and more precise measurements of their spatial offsets within these systems. This advancement will greatly improve the identification of dual, offset, and recoiling AGNs, thereby deepening our understanding of the intricate relationship between mergers, galaxy substructures, and AGN activities.

\begin{acknowledgements}
Based on observations with the NASA/ESA/CSA James Webb Space Telescope obtained from the Barbara A. Mikulski Archive at the Space Telescope Science Institute, which is operated by the Association of Universities for Research in Astronomy, Incorporated, under NASA contract NAS5-03127. Support for Program numbers JWST-GO-02057 and JWST-AR-03038 was provided through a grant from the STScI under NASA contract NAS5-03127. 
\end{acknowledgements}

\bibliography{sample631}{}
\bibliographystyle{aasjournal}

\end{document}